\documentclass[preprint,aps,preprintnumbers,amsmath,amssymb]{revtex4}
\usepackage{amsmath}
\usepackage{amsbsy}
\usepackage{amssymb}
\usepackage[dvips]{graphicx}
\usepackage{epsfig}
\usepackage{color}

\begin{document}

\title[]{Spectral analysis, chiral disorder and topological edge states manifestation in open non-Hermitian Su-Schrieffer-Heeger chains}

\author{B. Ostahie  and A. Aldea}

\address{National Institute of Materials Physics, 
077125  Bucharest-Magurele, Romania. }
\vspace{10pt}

\begin{abstract}
We investigate topological and disorder effects in non-Hermitian 
systems with chiral symmetry. The system under consideration consists 
in a finite Su-Schrieffer-Heeger chain to which two semi-infinite 
leads are attached. The system lacks the parity-time and time-reversal
symmetries and is appropriate for the study of quantum transport properties.
The complex energy spectrum is analyzed in terms of 
the chain-lead coupling and  chiral disorder strength, and 
shows substantial differences between  chains with even and odd 
number of sites. The mid-gap edge states acquire a finite lifetime
and are both of topological origin or generated by a strong 
coupling to the leads. The disorder induces coalescence of 
the topological eigenvalues, associated  with exceptional points 
and vanishing of the eigenfunction rigidity.
The electron transmission coefficient is approached in the 
Landauer formalism, and an analytical expression for the transmission 
in the range of topological states is obtained. Notably, the chiral disorder in this non-Hermitian system induces unitary conductance  
enhancement in the topological phase.
\end{abstract}

\maketitle

\section{Introduction}
{\color{black} Different non-Hermitian models have been developed mainly
for their exotic properties coming from the complex eigenvalues 
and biorthogonal eigenvectors, combined with different symmetry arguments.
For instance, much attention has been paid to non-Hermitian models
with parity-time ($\mathcal{PT}$) symmetry, which
exhibit both real and complex eigenvalues depending on 
internal parameters \cite{Bender,Bender1}. Such kind of  non-Hermitian  
Hamiltonians are used  for describing the  gain and loss  processes 
in optical lattices \cite{Nori,Pan,Poli,Song}.

There is a recent  trend to identify topological phases in
non-Hermitian systems. The question of defining bulk topological 
invariants and to find the correspondence to boundary states 
(the so-called bulk-boundary correspondence) is still a controversial 
discussion \cite{Yuce, Kunst, Yao, Xiong, Ueda, Lee, Leykam, Takane, Das, Torres}.
In parallel, the disordered topological matter is also of 
considerable interest \cite{Jain,Altland, Prodan}.

The  Su-Schrieffer-Heeger (SSH)  model is a platform commonly used 
for developing new topological concepts. Looking for new physics, 
the  Hermitian model was  extended by adding next-nearest-neighbor-hopping 
\cite{Platero}, disorder \cite{Prodan}, or  expanding the 
unit cell \cite{Xie}. In this paper, we look for the  properties 
of the topological phase in  {\it open mesoscopic} (finite-size) 
SSH systems. In this case, the non-Hermiticity is induced by the 
coupling  to external reservoirs, and  describes the electron 
transport problem. More explicitly, while the confined (finite) 
system is described by a Hermitian Hamiltonian, the open system 
obtained by attaching semi-infinite wires can be described by an  
effective Hamiltonian, which turns out to be non-Hermitian 
\cite{Rotter, Moldoveanu, ONA}. As for the symmetries, the non-Hermitian 
Hamiltonian related to the open system lacks the time-reversal 
$\mathcal{T}$ and $\mathcal{PT}$ (its eigenvalues being always complex), 
but keeps the chiral symmetry $\mathcal{C}$ of the SSH chain.

In the present study, we demonstrate  effects arising from 
non-Hermiticity and disorder, the purpose of the paper being  
twofold: i) to look for the fate of the mid-gap topological edge states 
when the coupling (hybridization) to the leads is introduced gradually, 
and ii) to study the competition between the  coupling parameter
(which controls the non-Hermiticity) and the disorder strength.
We deliberately analyze the  non-diagonal disorder
(which affects the chain bonds) as it  preserves the chiral 
symmetry of the SSH system and has the advantage of allowing 
an analytical solution.

Spectral and quantum transport quantities, describing  the response 
of the  topological states to the chain-lead coupling and  
chiral disorder, are calculated analytically and numerically. 
We deduce the exact analytical expression (22) of the electron  
transmission on the topological states, which shows a non-monotonic 
dependence on the coupling parameter and also a unitary peak. 
This behavior is explained in terms of density of states
and is supported by the dwell time and the reflected flux delay.

The dependence of the transport properties on the disorder strength is 
nontrivial  and allows for a disorder-induced conductance  in the case 
of sufficiently strong coupling. The origin of this effect consists 
in the delocalization process  of the edge states generated  by 
the chiral disorder. It is to mention that the unitary transmission 
is conserved even in the disordered case.

We also find that, under the condition of strong  hybridization, 
mid-gap edge states  may occur even in the non-topological  range of 
parameters. Another finding of interest is the level coalescence 
in the topological and non-topological phase, which  can be 
induced by the chiral disorder, and depends on the chain length 
and the chain-lead coupling.

The paper is structured as follows: In  Section II, we discuss 
the spectral properties of the topological phase for  ordered 
and disordered finite SSH chains. The spectral properties of 
the non-Hermitian Hamiltonian in topological/non-topological phase,
including  the coalescence problem and the phase rigidity,
are evidentiated in Section III. In Section IV, the quantum transport 
on the topological states (transmission coefficient, dwell time 
and reflected flux delay) is analyzed analytically and numerically. 
The section also highlights  the chiral disorder effects on the density 
of states and transport properties. The results are summarized 
in the last section.

\section{Spectral  properties of the finite
SSH-chain with chiral disorder}
The SSH model consists in  a bipartite one-dimensional diatomic chain
with alternating hopping parameters $t_1$ and $t_2$, which connect
the sites of type $A$ and $B$. The model is described by the
following tight-binding spinless Hermitian Hamiltonian:
\begin{equation}
H=t_1\sum_i c_{2i-1}^{\dag}c_{2i}+
t_2\sum_i c^{\dag}_{2i}c_{2i+1}
+H.c..
\end{equation}
As we shall see,  there are significant differences in the
spectral and transport properties of the finite chains with {\it even}
or {\it odd} number of sites. It is to remind that in the case
of infinite chains, by imposing periodic boundary conditions,
the Hamiltonian (1) can be rewritten in the momentum representation as:
\begin{equation}
\hskip1.0cm
H=\sum_{\vec{k}}
\left(\begin{array}{cc}
a^{\dagger}_k~ & b^{\dagger}_k
\end{array}\right)
H(k)
\left(\begin{array}{c}
    a_k \\
    b_k\\
\end{array}\right),~~
H(k)=(t_1+t_2cosk)\sigma_x+t_2 sink\sigma_y.~~~~~~~~~~
\end{equation}
where $\sigma_x,\sigma_y$ are Pauli matrices.
The above Hamiltonian shows chiral symmetry  $\{\sigma_z, H(k)\}=0$,
but lacks both the parity $\mathcal{P}$ and time-reversal
$\mathcal{T}$ symmetries. That is, the  SSH model  belongs to
the AIII class  \cite{Zirnbauer,Ryu} and may show a topological
transition controlled by the ratio $t_2/t_1$ and specified by a
winding number $\nu$.

In what concerns the {\it finite} chains, one may expect
that the chain ends play a relevant role for the spectral and
topological properties. Let us denote  by $N_A$ and $N_B$,
the number of A and B atoms in the finite chain, respectively.
Assuming that  the chain  begins with A and ends also  with A,
the total number of lattice sites $N=N_A+N_B$ is odd, and $N_A-N_B=1$.
Then, a theorem valid for {\it bipartite} lattices  states that
the energy spectrum contains {\it at least} one eigenvalue in
the middle of the gap  at $E=0$  \cite{Lieb, Mielke}.
On the other hand, if the chain begins with A and ends with B, i.e.,
the total number of sites $N$ is even, one has  $N_A-N_B=0$,
and  the minimum number of allowed states
at the midgap equals zero.

In what concerns the topological properties, a guess can be done
in advance. Since for N=odd the number of $t_1$-bonds and of
$t_2$-bonds are equal, the ratio $t_2/t_1$ becomes meaningless
as a  criterion for a  topological transition, so that we do not
expect  topological properties for SSH chains with an odd number of sites.

On the contrary, for N=even, as  consequence of the bulk-edge
correspondence, two topological edge states should appear if
$t_2/t_1>1$, but they cannot be located strictly at  zero energy,
as this is forbidden by the before-mentioned theorem.
The emergence of the topological states can be visualized
by reducing gradually the hopping parameter
$t_{N,1}$ between the last  and the first site of the chain
with periodic boundary condition.
The decrease of $t_{N,1}$ breaks the periodicity and gives rise to
a pair of states in the gap, which become quasi-degenerate in
the middle of the gap as shown in Fig.1.
The eigenvalues of several disordered configurations of the same chain
are also shown in Fig.1 (blue points) in order to prove the
robustness of the topological states created near $t_{1N}=0$
(in the domain encircled by the black curve).
\begin{figure}[htbp]
\centering
\includegraphics[scale=0.3, angle=-90]{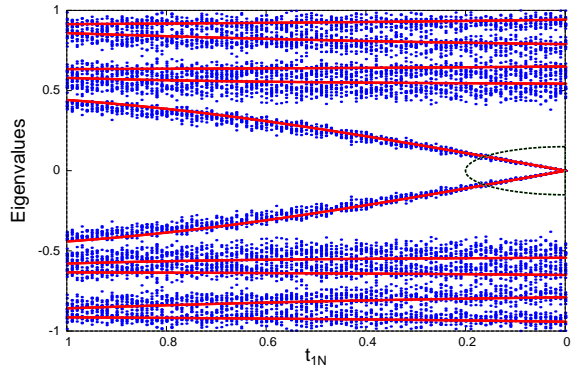}
\caption{The crossover  of the energy spectrum of the  finite chain from
{\it periodic}  ($t_{1N}=1$) to {\it open} ($t_{1N}=0$) boundary condition.
Eigenvalues of ten disordered configurations  are
superimposed over the spectrum of the ordered chain (shown in blue and red,
respectively). The chain length is $N=40$, disorder strength
is $W=0.7$, and  $t_2/t_1=1.5$.}
\end{figure}
In what follows, we approach the question of the quantum states located
in the middle of the gap in the case of {\it finite} SSH chains with an
even number of sites, the only ones which are interesting from the
topological point of view.  Looking  for the wave functions as:
\begin{equation}
|\Psi\rangle=\sum_{n=1}^{N_c}(\Psi^A_n a^{\dag}_n|0\rangle+
\Psi^B_n b^{\dag}_n|0\rangle),
\end{equation}
the following  recurrence relations are obtained:
\begin{eqnarray}
\hskip2.5cm
t_1\Psi_n^A+t_2\Psi_{n+1}^A=E\Psi_n^B \nonumber\\
\hskip2.5cm
t_1\Psi_n^B+t_2\Psi_{n-1}^B=E\Psi_n^A.
\end{eqnarray}
As consequence of the inversion symmetry shown by the Hamiltonian
in the case of  even chains, one finds $|\Psi^A_n|=|\Psi^B_{N_c-n+1}|$.
As  the Hamiltonian anticommutes with the chiral operator
\begin{equation}
\mathcal{C}=\sum_{n=1}^{N_c}(a_n^{\dag} a_n-b_n^{\dag} b_n),
\end{equation}
it is trivial to prove the electron-hole symmetry of the energy spectrum.

The question of finding the two quasi-degenerate energies in the middle
of the spectrum will be approached perturbatively.  In this spirit,
let us accept for the beginning that  $E=0$ is an eigenvalue
and calculate  the corresponding  $\Psi^A_n and  \Psi^B_n$.
The recurrence  relations  (4) provide the following behavior of
the coefficients:
\begin{eqnarray}
\hskip2.5cm
\Psi^A_n=(-t_1/t_2)^{n-1}\Psi^A_1~~\nonumber \\
\hskip2.5cm
\Psi^B_n=(-t_1/t_2)^{N_c-n}\Psi^B_{N_c}.
\end{eqnarray}
In the topological phase ($t_2/t_1>1$), $\Psi^A_n$
shows its maximum  at $n=1$, while $\Psi^B_n$ reaches the  maximum
value at the other end of the chain $n=N_c$. It turns out that the
functions $|\Psi^A>=\sum_n\Psi^A_n a^{\dag}_n|0>$ and
$|\Psi^B>=\sum_n\Psi^B_n b^{\dag}_n|0>$ represent two orthogonal
edge states, each one being localized at another end of the chain.
The  penetration length $\lambda$, defined as $|\Psi^A_n/\Psi^A_1|=
exp\{-(n-1)/\lambda\}$, is given by:
\begin{equation}
\lambda^{-1}=\frac{1}{n-1}ln|\Psi^A_n/\Psi^A_1|=-ln(t_1/t_2),
\end{equation}
and indicates that the strong localization of the edge states
occurs for small ratios $t_1/t_2$. Obviously, the functions
$\{|\Psi^A>,|\Psi^B>\}$ are only approximate eigenvectors
of the finite chain Hamiltonian, corresponding to the approximate
eigenvalue $E=0$. The actual eigenvectors can be written as
their superposition, so that
the secular equation  gives the eigenvalues:
\begin{equation}
\hskip-0.14cm
E_{\pm}=\pm|<\Psi^A| H \Psi^B>|=\pm t_2(t_1/t_2)^{N_c}
|\Psi^{A*}_1 \Psi^B_{N_c}|.
\end{equation}
In conclusion, the  above analytical calculation provides the splitting
$\Delta=E_{+}-E_{-}$ between the energies of the two topological
edge states and  proves  the exponential decay of the splitting
in the limit of  long chains.

In what follows we discuss the disorder effects on
the penetration length $\lambda$ of
the topological edge states is affected by disorder.
we take into account  the  off-diagonal disorder obtained
by considering the hopping parameter $t_2$ as a random variable,
while $t_1$ is chosen as the energy unit ($t_1=1$).
This is a {\it chiral-type} disorder in the sense that the
Hamiltonian describing the disordered system
preserves the anticommutation relation with the chiral operator
$\mathcal{C}$ defined in (5), and, consequently, preserves
the electron-hole spectral symmetry. To be specific, the
hopping parameter $t_2$ is uniformly distributed in the range
$[t_2-\frac{W}{2},t_2+\frac{W}{2}]$, where $W$ measures
the disorder strength.
Since  the parameter $t_2$ becomes a random variable,
Eq.(6) should  be replaced by:
\begin{equation}
\Psi^A_n=\Pi_{i=1}^{n-1}(-t_1/t_2^{(i)})~\Psi^A_1,
\end{equation}
while  the penetration length  (7) should be redefined
as the configurational average over all  disorder realizations.
Assuming a disorder strength  $W<2t_2$ and $t_1=1$, one obtains:
\begin{equation}
\hskip-0.15cm
\frac{1}{\lambda}=\frac{-1}{n-1}<ln|\Psi^A_n/\Psi^A_1|>
=\int_{-1/2}^{1/2}ln(t_2+Wx)dx,
\end{equation}
which yields the following analytical expression:
\begin{equation}
\frac{1}{\lambda}=\frac{1}{W}[-W+(t_2+\frac{W}{2})ln(t_2+\frac{W}{2})
-(t_2-\frac{W}{2})ln(t_2-\frac{W}{2})].
\end{equation}
The  dependence of $\lambda$  on the disorder strength $W$ is shown in
Fig.2. One remarks the localization effect induced by the increase of $t_2$.
However, more interesting is the delocalization process of the
topological edge  states expressed by  the increasing penetration
length $\lambda$, which takes place with increasing disorder.
This result is in line with the finding in \cite{Prodan},
where, for a generic one-dimensional (1D) disordered model,
the topological phase ($\nu=1$) persists up to a
given degree of disorder, where a localization-delocalization
transition occurs.
\begin{figure}[htbp]
\centering
\includegraphics[scale=0.3, angle =-90]{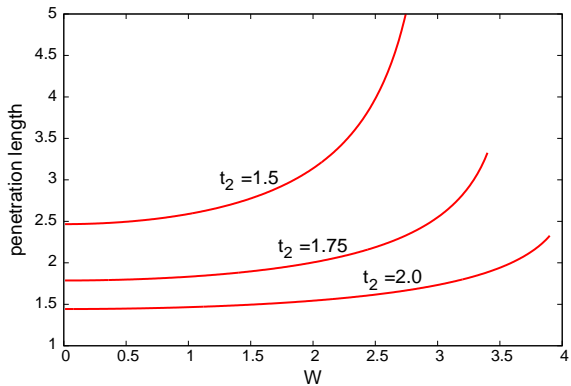}
\caption{The penetration length $\lambda$ as function of the disorder
strength $W$ for several values of the hopping parameter $t_2$.
The increase of $\lambda$ indicates a delocalization process of
the edge states.}
\end{figure}

\section{Spectral properties of the open SSH-chain: 
exceptional points and the rigidity of the wave function}
The open system consists in a finite SSH chain to which 
two semi-infinite leads (probes) are attached at the sites n=1 and n=N, 
via the  coupling parameter $\tau_c$ (see Fig.3). The leads are 
one-dimensional and described by an energy spectrum $E_k=2t_L cos k$.
When coupled to leads, the SSH chain can be described by an 
{\it effective} Hamiltonian, which turns out to be {\it non}-Hermitian,
with  broken time-reversal  symmetry:
\begin{equation}
H_{eff}=H + {\frac{\tau_c^2}{t_L}}e^{ik} (c_{1}^{\dag}c_{1}+c^{\dag}_{N}c_{N})
\end{equation}
It is important to notice that $H_{eff}$ depends, via the parameter $k$, 
on the Fermi energy $E_F$ of the electrons in the leads.
Throughout the paper, this energy will be chosen $E_F=0$, 
corresponding the $k=\pi/2$ in (12). 
At the same time, the hopping parameter $t_1$ present in $H$, 
is chosen as the energy unit ($t_1=1$), so that the problem is defined 
in the  parameter space $\{t_2,\tau_c\}$. As already discussed, 
the length $N$ of the SSH-chain is also an  important quantity.

The deduction of Eq.(12) is rather simple and is based on a 
projection technique (see, for instance, the Green function approach in 
\cite{ONA}). 

\begin{figure}[htbp]
\centering
\includegraphics[scale=0.8]{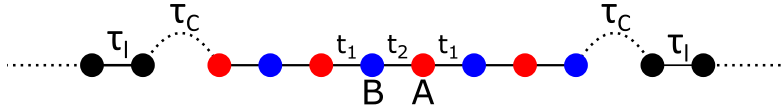}
\caption{
Schematic representation of the SSH chain, where A (red) and B(blue) 
represent the two atoms in the unit cell, $t_1$ connects the atoms in the 
same cell, and $t_2$ connects the atoms in adjacent cells.  
The first and the last site  are connected to semi-infinite leads 
via the  coupling parameter $\tau_c$, while $\tau_l$ is the hopping 
amplitude in the leads.}
\end{figure}

In order to obtain the complex eigenvalues of (12), one has 
to diagonalize the following matrix (written below for an even 
number of sites):
\begin{equation}
{\bf H_{eff}} =
\begin{pmatrix}
i\tau_c^2/t_L & t_1 & 0 & ... & 0 \\
t_1 & 0 & t_2 & ... & 0 \\
... & ... & ... & ... & ... \\
0 & ... & t_2 & 0 & t_1 \\
0 & ... & 0 & t_1 & i\tau_c^2/t_L
\end{pmatrix}.
\end{equation}

Before discussing the complex spectrum of longer chains, it is helpful to 
analyze two short chains, with N=3 and N=4, which allow analytical 
solutions. These examples make clear the spectral differences between the  
even and odd chains, mainly in the context of the so-called 
{\it exceptional points}. For $N=3$, the three eigenvalues read:
\begin{eqnarray}
\omega_0=i\tau^2_c/t_L,~~~~~~~~~~~~~~~~~~~~~~~~~~~~~~~~~~~~~~~~~\nonumber \\
\omega_{\pm}=
\frac{1}{2}\big(i\tau^2_c/t_L\pm \sqrt{(i\tau^2_c/t_L)^2+4(t_1^2+t_2^2)}~\big).
\end{eqnarray}
For $N=4$, the four eigenvalues read:
\begin{eqnarray}
\omega_{1,\pm}=\frac{1}{2}\big(i\tau^2_c/t_L+
t_2\pm \sqrt{(i\tau^2_c/t_L+t_2)^2+4t_1^2)}~\big),
\nonumber\\
\omega_{2,\pm}=\frac{1}{2}\big(i\tau^2_c/t_L-
t_2\pm \sqrt{(i\tau^2_c/t_L-t_2)^2+4t_1^2)}~\big).
\end{eqnarray}
An  exceptional point (EP) is defined by the degeneracy  
of two eigenvalues of the non-Hermitian Hamiltonian, 
which comes out at a given value of the coupling $\tau_c$.
At such a point, one occurs the coalescence at $Re E=0$ of the real parts 
of the two eigenvalues and also the bifurcation of the imaginary parts. 

One notices from Eq.(14) that, in the odd case N=3, 
the condition $\omega_+=\omega_-$  is fulfilled  when the quantity 
under the radical vanishes, i.e. at $\tau_c^2=2t_L\sqrt{t_1^2+t_2^2}$.
One  observes also that  the imaginary part of the energies
$\omega_{\pm}$ is a nonanalytical function of $\tau_c$, as all 
derivatives diverge at EP.

On the contrary, in the case N=4 described by Eq.(15), 
the necessary condition for an EP, which in this case reads as
$\omega_{1-}=\omega_{2+}$, cannot be fulfilled (except for $t_2=0$,  
meaning in fact a broken chain). 


\begin{figure}[htbp]
\centering
\includegraphics[scale=1.5]{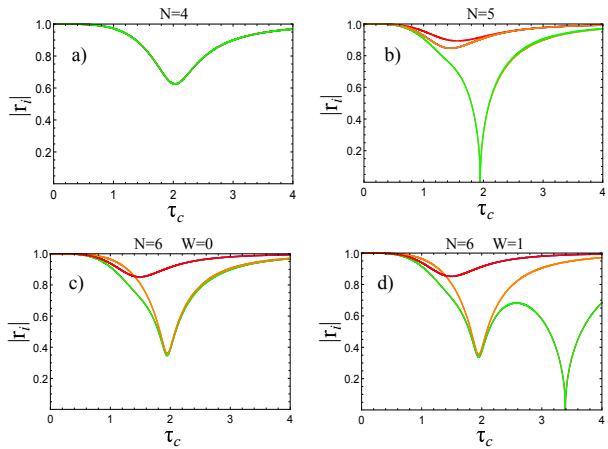}
\caption{Phase rigidity in the non-topological case ($t_2/t_1=0.5$)
for ordered/disordered chains with even/odd number of sites as  
function of the coupling strength. Panels a) and c) show that  
for ordered even chains of length $N=4,6$ the EP is missing,  
however it appears for the ordered odd chain with $N=5$ (panel b). 
The significant result is that the chiral disorder induces EP 
in an even chain, too (see the green line in panel d).}
\end{figure}

The question of exceptional points can be discussed also in terms of
eigenfunction {\it phase rigidity}, defined as:
\begin{equation}
r_i=\frac{|<\Psi_i^L|\Psi_i^R>|}{<\Psi_i^R|\Psi_i^R>},
\end{equation}
where $<\Psi_i^L|$ and  $|\Psi_i^R>$ are the left and right biorthogonal
eigenvectors of the  non-Hermitian Hamiltonian, and  $i$ is the 
eigenvector index. The rigidity measures the degree of biortogonality
of the eigenfunctions and depends on the strength of the coupling
with the infinite leads. Obviously, for the Hermitian case $\tau_c=0$ one has 
$<\Psi_i^L|=<\Psi_i^R|$ such that $r_i=1$. With increasing $\tau_c$ the
rigidity decreases, and one proves \cite{Rotter} that the coalescenting
states show $r=0$ at the EP. This is the case in Fig.4b for an
ordered chain of length $N=5$.
As a counterexample, Fig.4a and Fig.4c show a 
non-vanishing phase rigidity for chains with  an even number of  
sites ($N=4$ and $N=6$, respectively) and under non-topological
condition $t_2/t_1<1$.
Nevertheless, the  presence of  rigidity dips in these figures 
indicates  a {\it quasi}-degeneracy of the real parts, and a 
smooth branching of the imaginary parts of the  involved eigenvalues. 
One may suspect that in the limit of $N\rightarrow\infty$ the rigidity 
$r_i\rightarrow 0$.

The  analytical formulas for N=3 and N=4  and also the numerical results 
for longer chains suggest that EPs are exhibited only by SSH chains with 
an odd number of sites. The distinction between the  odd and even chains 
consists in the absence, and respectively, the presence of the   
spatial inversion symmetry.   Apparently, the  presence of this symmetry
prevents the appearance of  exceptional points in the open SSH chains 
with even number of sites. We  have not a proof of that, 
nevertheless the study  of disordered chains gives support to  this argument.
Indeed, the chiral disorder introduced into chains with N=even  
breaks  the previous inversion symmetry,
and, at the same time, gives rise  
to a vanishing phase rigidity. This result is obvious by comparing Fig.4c and Fig.4d.

\begin{figure}[htbp]
\centering
\includegraphics[scale=1.8]{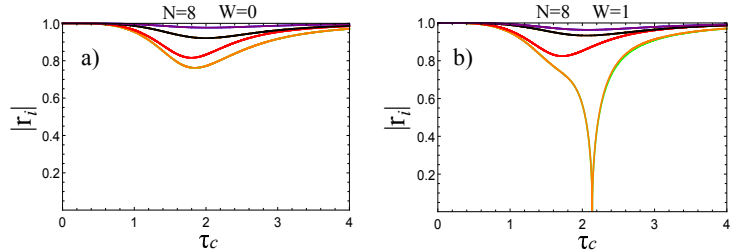}
\caption{Phase rigidity in the topological case ($t_2/t_1=1.5$) as function of the coupling strength. The states  showing  the vanishing rigidity in the presence of disorder ($W=1$) are the topological ones.}
\end{figure}

The eigenfunction phase rigidity induced  by the non-Hermitian 
perturbation occurs also in the topological phase ($t_2/t_1>1$), 
as shown in  Fig.5. The vanishing rigidity is  due again to the 
chiral disorder, and it is exhibited by the topological eigenstates.

The way  the parameter $\tau_c$ controls this effect can be followed by looking at the flow of the eigenvalues in the complex plane \{Im E, Re E\} when $\tau_c$ is varied (see Fig.6).
At $\tau_c=0$, the eigenvalues  are real and  the splitting between the two topological levels  is that one given  by Eq.(8).
An analytic perturbative approach shows readily that for small $\tau_c$ the real parts  of the  topological eigenvalues attract each other. 
By numerical calculation, one finds that also for larger coupling
the real parts keep getting closer, while the imaginary part increases.
Finally, there is a value of $\tau_c$ where the  real parts of the 
two eigenvalues coalesce at $Re E=0$, while the imaginary parts 
continue to increase.

Strictly specking, only the disordered chain (blue curve in Fig.6)
shows a rigorous coalescence, which corroborates the exceptional point 
shown by the phase rigidity in Fig.5b. In what concerns the ordered 
chain (red curve in Fig.6), the coalescence is only approximate, 
in agreement with the above discussion.

\begin{figure}[htbp]
\centering
\includegraphics[angle=-90, scale=0.3]{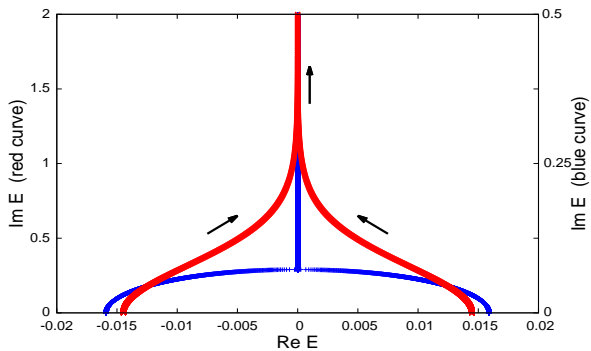}
\caption{The flow in the complex plain of the eigenvalues 
corresponding to the two topological edge states, when the 
chain-lead coupling $\tau_c$ is increased continuously. 
The ordered chain is described by the red curve, while a disordered 
chain with $W=1$ is described by the blue one. The arrows indicate 
the increasing $\tau_c$. (other parameters: $t_2/t_1=1.5, N=20$).}
\end{figure}
Taking into account  that  for the infinite SSH chain the ratio  
$t_2/t_1$ dictates the presence of the topological  phase,
it becomes of interest to study the role of this parameter  also 
in the case of  open finite chains. 
In what follows, 
we illustrate how  the external leads 
affects the midgap edge modes already developed  in the confined  
finite chain (and discussed in Section II).  
We also detect a new type of midgap edge states in the non-topological 
range ($t_2/t_1<1$), which are generated by  the strong coupling 
to the leads.

The real and imaginary part of the  energy spectrum of the  {\it open} 
SSH chain of length N=20 is shown in Fig.7 as function of the 
parameter $t_2/t_1$, for different values of the chain-lead 
coupling $\tau_c$.
The  weak coupling case ($\tau_c=0.1$) shows in Fig.7a that the real 
part of two states detach themselves gradually from bands with 
increasing ratio $t_2/t_1$ and  merge continuously into two 
quasi-degenerate midgap edge states. At the same time, the eigenvalues 
acquire a large imaginary part shown in Fig.7b, which denotes  a short 
lifetime (defined as $\tau=\hbar/Im E$) of the topological edge states 
in the presence of the leads. This behavior of the complex spectrum 
at weak coupling is equivalent to the transition from  the insulating 
to topological phase for an infinite chain (without leads), 
which however occurs sharply at $t_2/t_1=1$.  
\begin{figure}[htbp]
\centering
\includegraphics[angle=-90, scale=0.2]{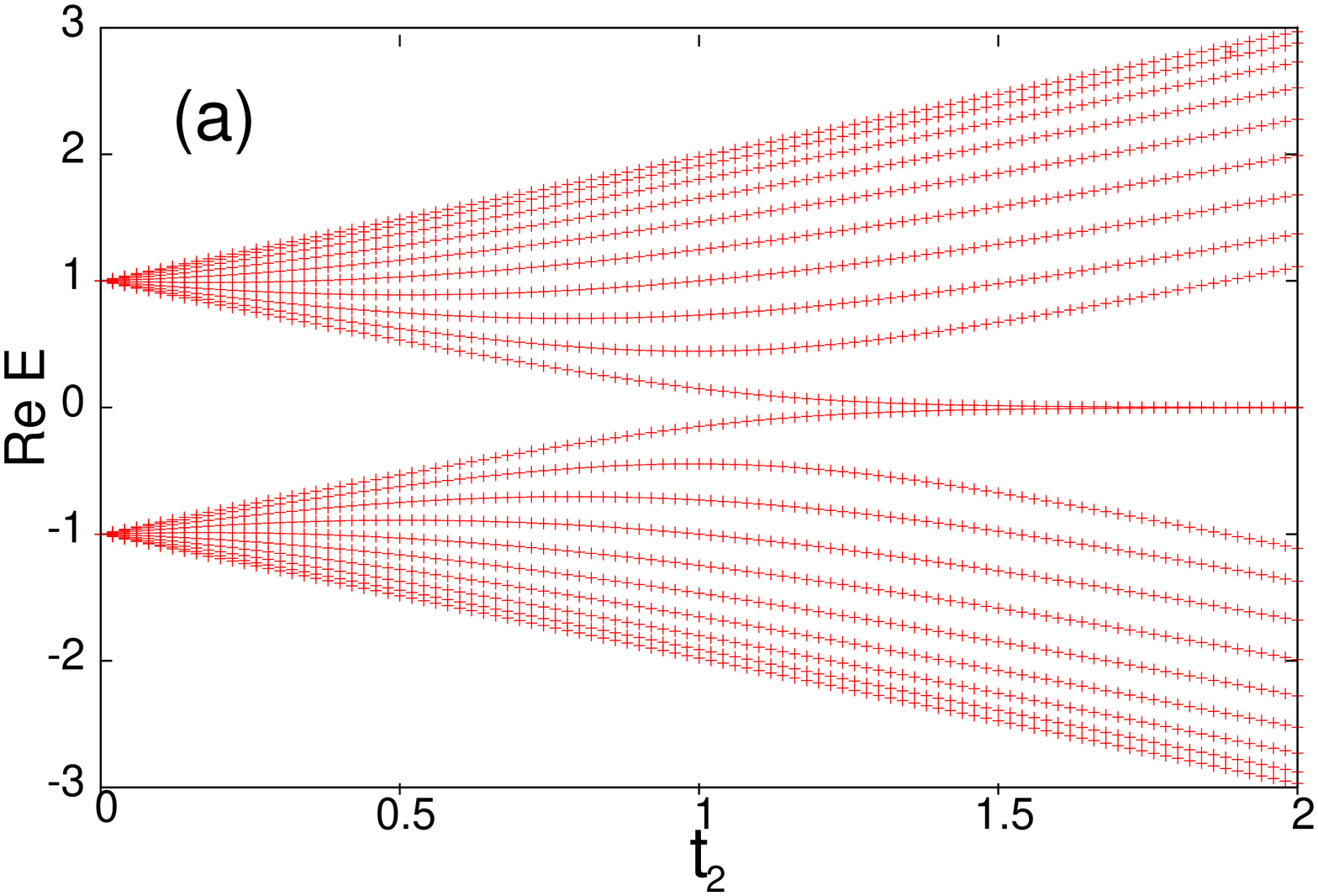}
\includegraphics[angle=-90, scale=0.2]{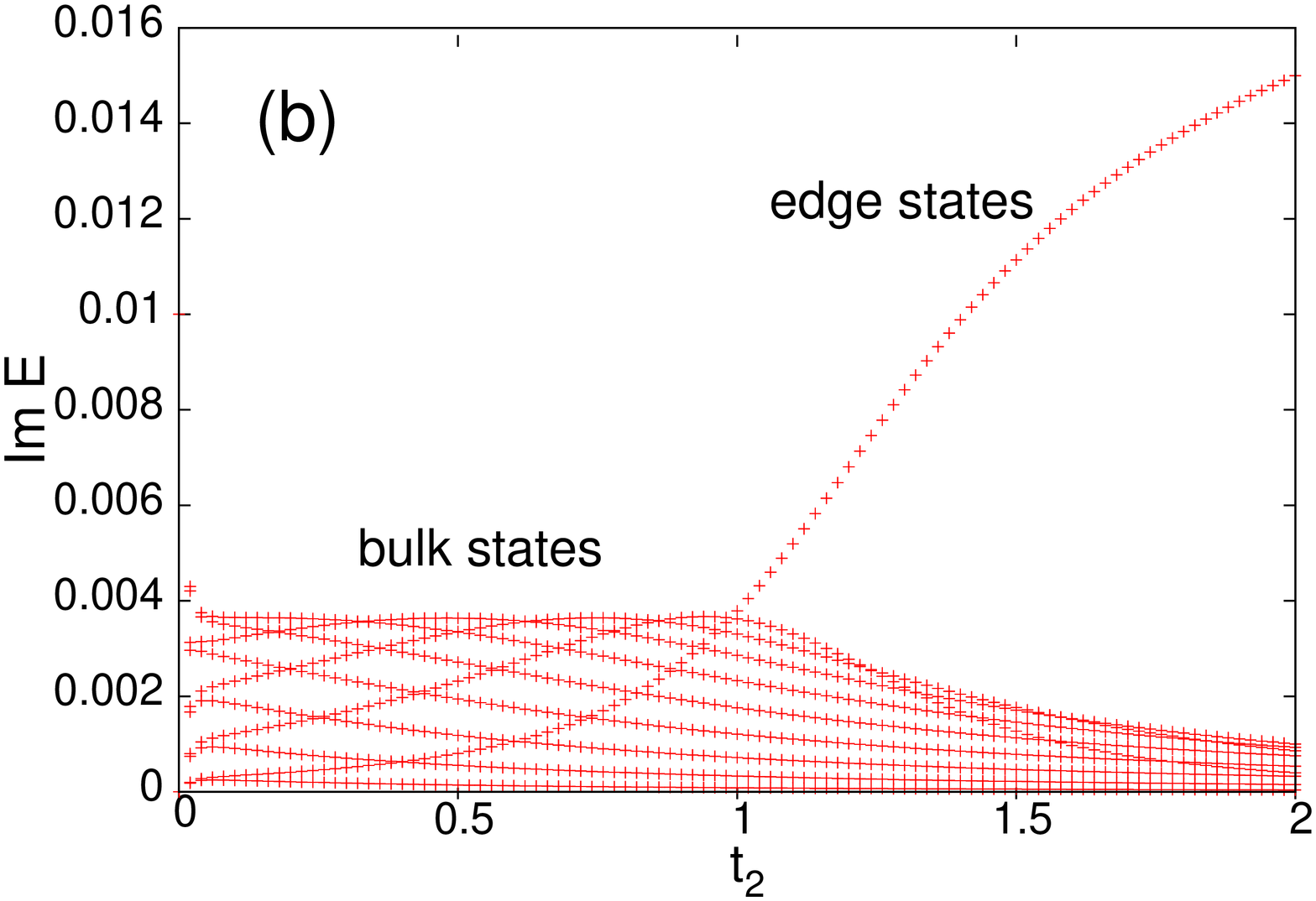}\\
\includegraphics[angle=-90, scale=0.2]{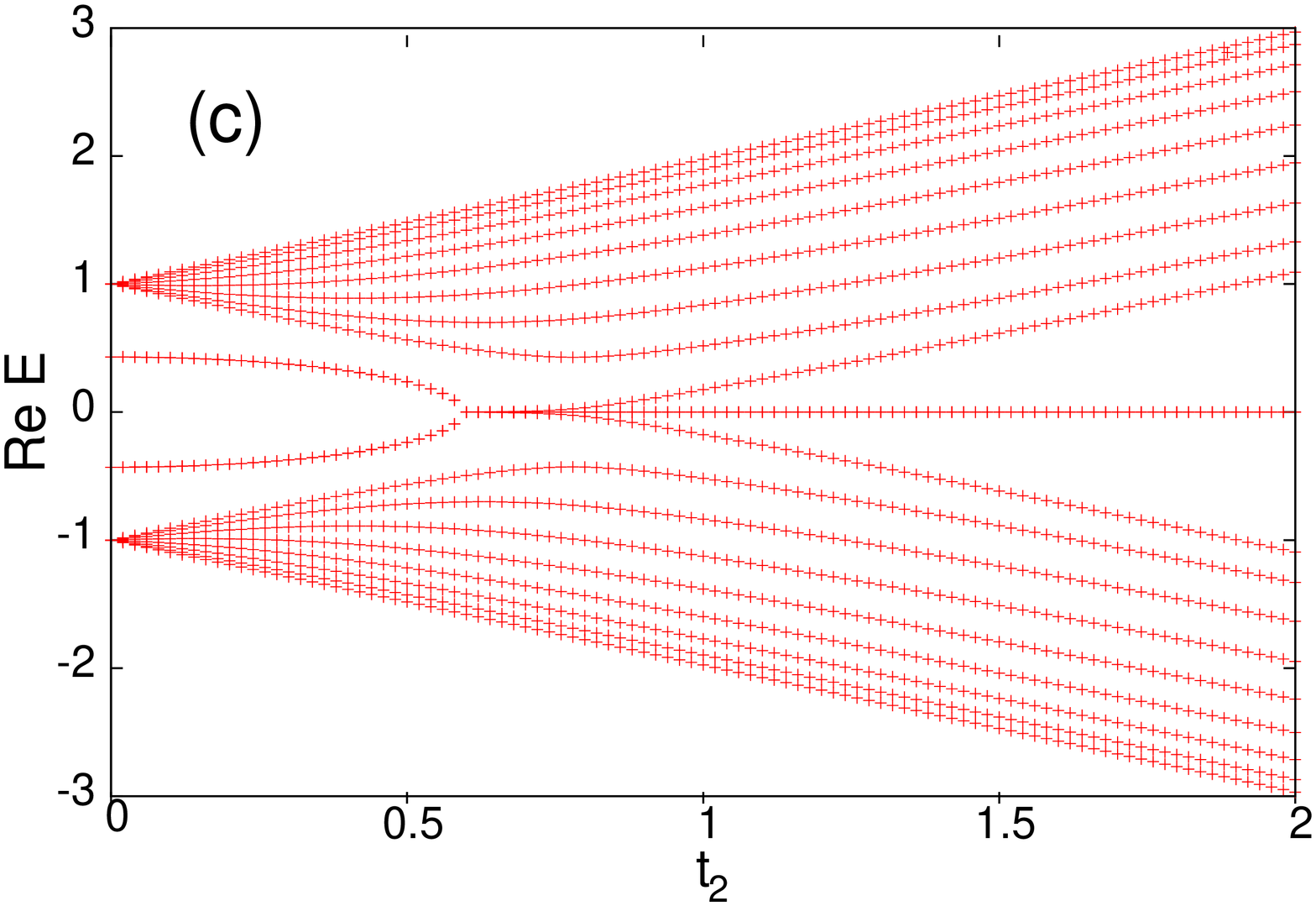}
\includegraphics[angle=-90, scale=0.2]{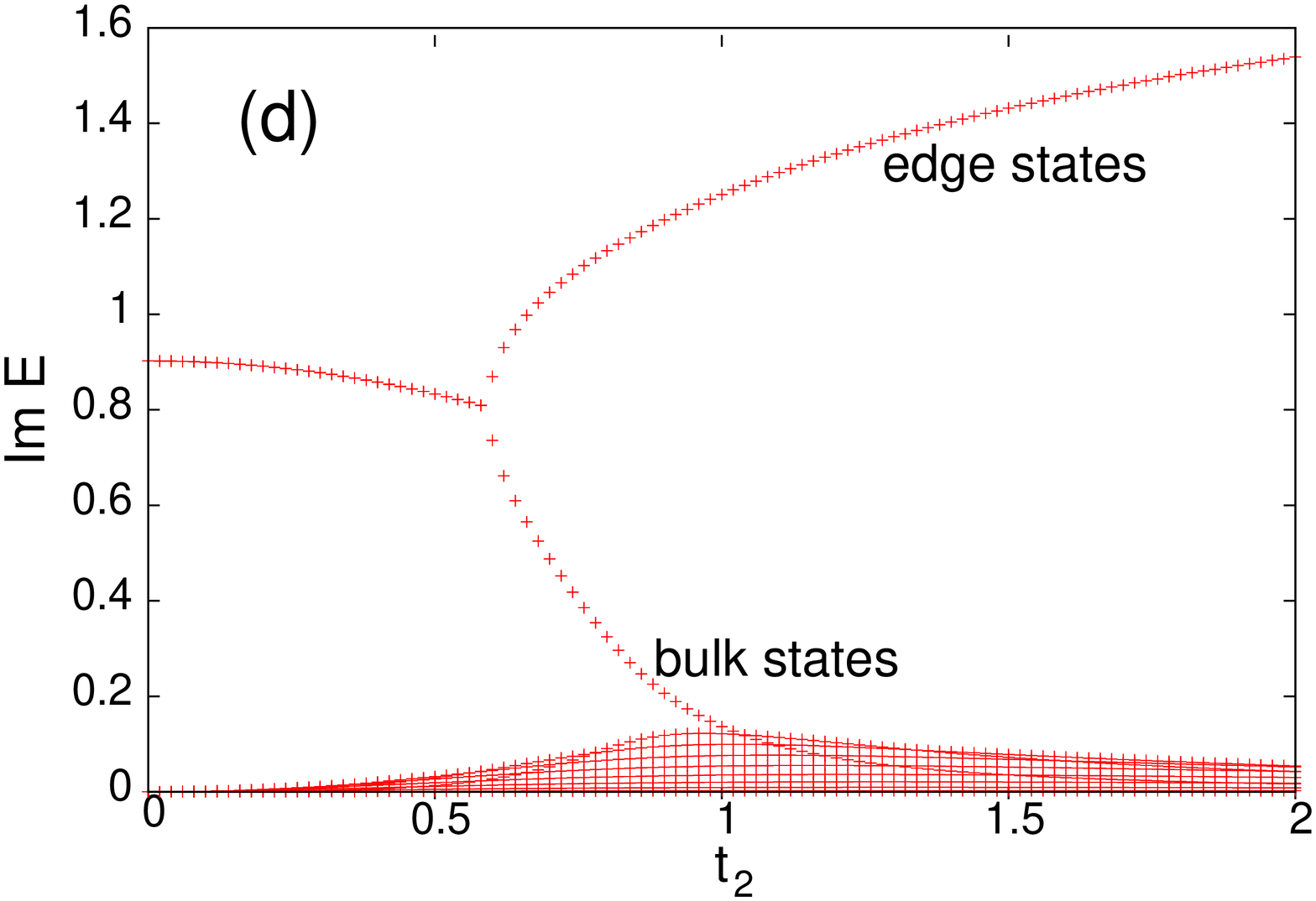}\\
\includegraphics[angle=-90, scale=0.2]{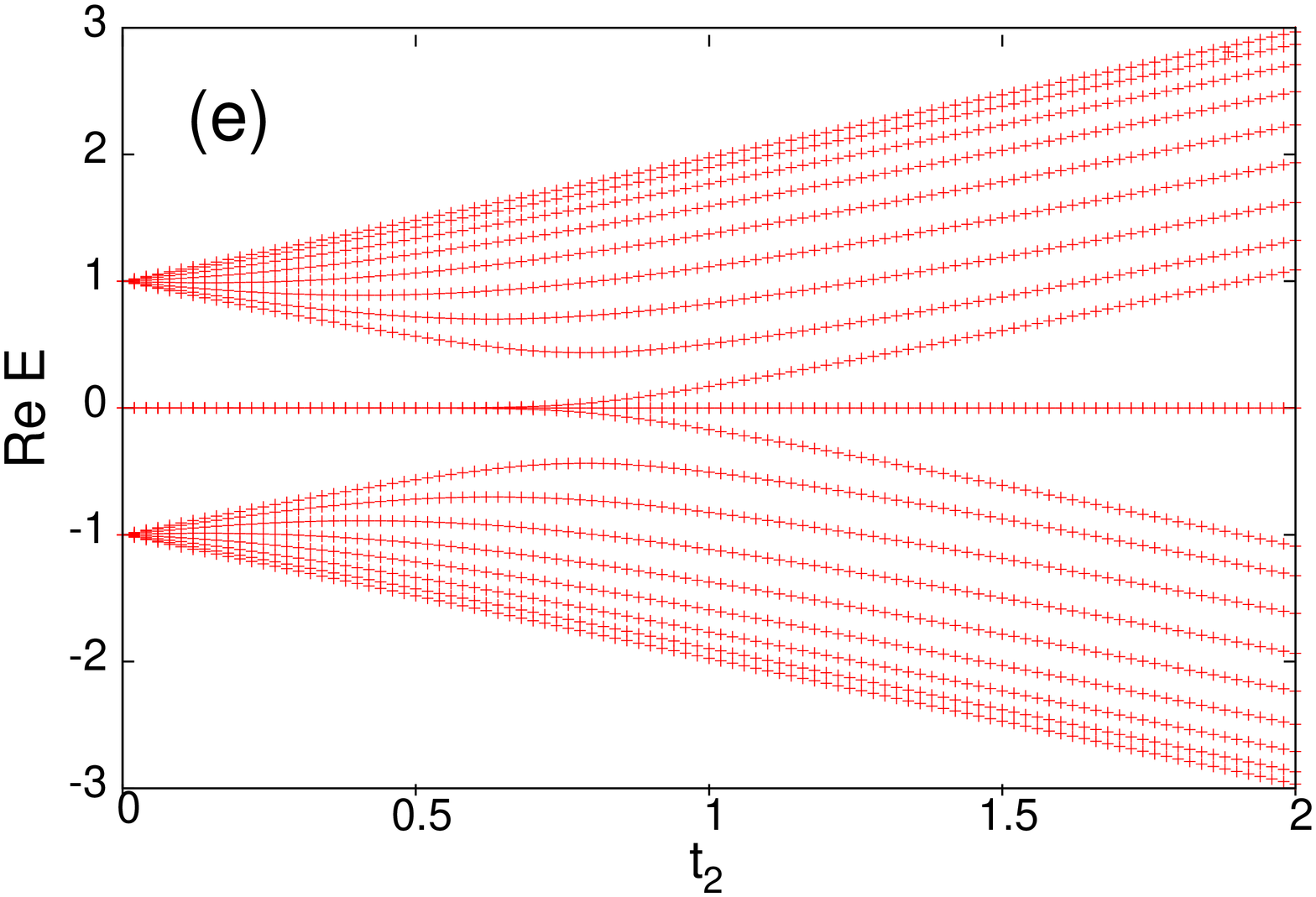}
\includegraphics[angle=-90, scale=0.2]{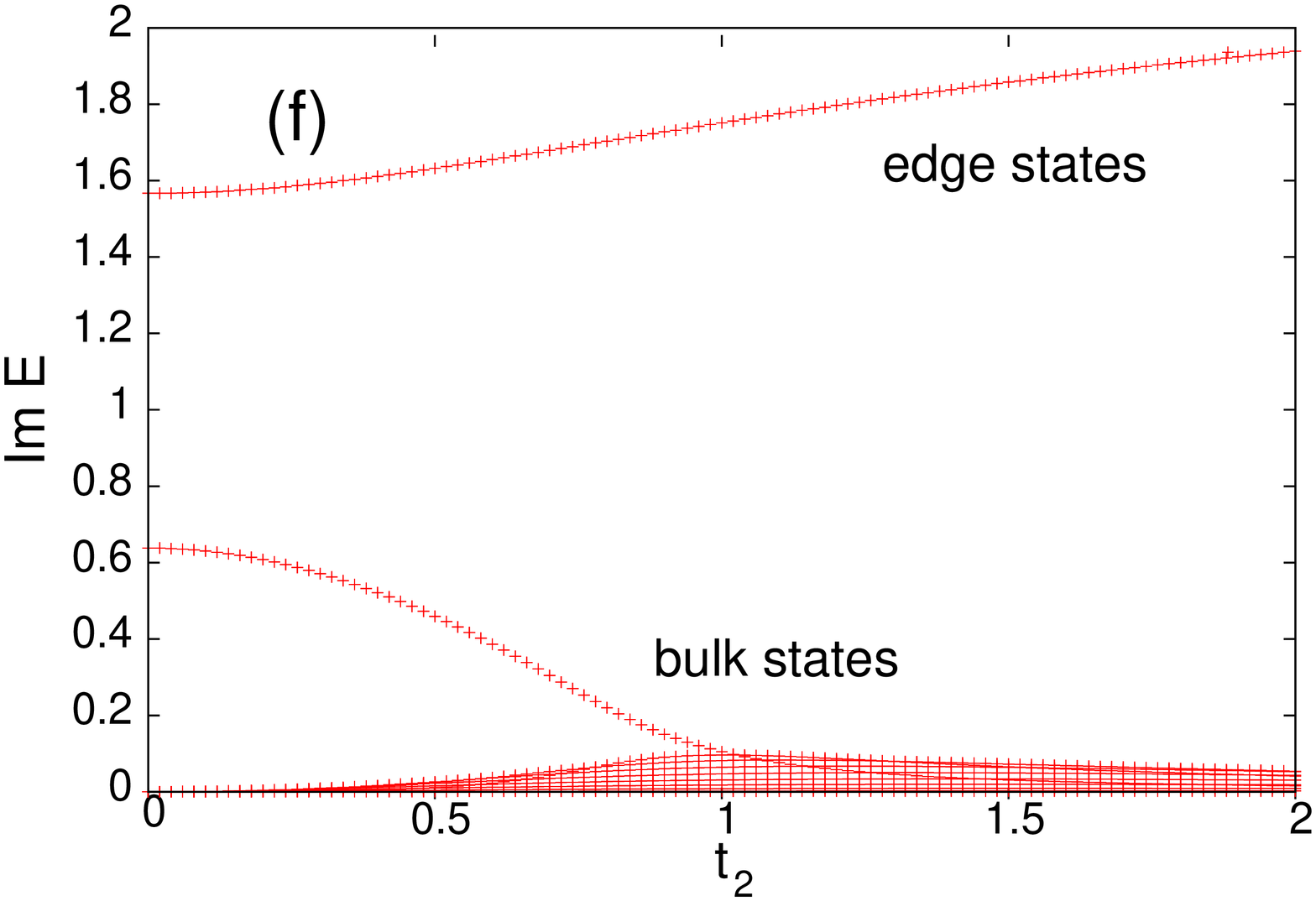}
\caption{
Real and imaginary parts of the eigenvalues of the effective 
Hamiltonian  (12) as  function of the hopping amplitude $t_2$, 
for an open chain with $N=20$, at different chain-lead couplings: 
$\tau_c=0.1$ (panels (a) and (b)), $\tau_c=1.9$ (panels (c) and (d)),
$\tau_c=2.1$ (panels (e) and (f)).
The imaginary part indicates that the edges states have a much shorter 
lifetime $\tau=\hbar/ImE$ than the bulk states. }
\end{figure}

An increased coupling  reveals new physics as illustrated in Fig.7c 
and Fig.7d. This time, two  levels  detach themselves  from each band 
at small $t_2$, next they become quasi-degenerate at $Re E=0$,
and,  finally, they split at large $t_2$ into two  states that return 
to the bands, and the two topological mid-gap edge states.
This level separation exhibited by  the real part of the spectrum 
is accompanied by a branching of the corresponding imaginary parts, 
which show large values (i.e., small lifetime) for  the edge modes,  
but a decreasing imaginary energy (i.e., increasing lifetime)  
of the states returning to the bands.

Fig.7e and Fig.7f describe the case of an  even stronger coupling 
($\tau_c=2.1$). This time, four mid-gap edge states are generated, 
which, with increasing $t_2$, evolve again into two longlife band 
states and two topological edge states with a small lifetime.
The hint for understanding why there are four edge states at small $t_2$ 
and strong $\tau_c$ comes from the formation of A-B dimers at each  
end of the chain, which sum up a total number of four quantum states. 
Strictly at $t_2=0$, the eigenvalues corresponding to such a dimer 
can be obtained immediately as:
\begin{equation}
E=\frac{1}{2}\big(\frac{i\tau_c^2}{t_L}
\pm \sqrt{-\frac{\tau_c^4}{t_L^2}+4t_1^2}~\big),
\end{equation}
and it is obvious that for sufficiently big $\tau_c$, the eigenvalues 
are purely imaginary, i.e., the states become of midgap-type.

The above examination of the spectral properties emphasizes that the coupling 
to the leads assigns a finite lifetime to the  topological edge states, 
which is much shorter than that one of  the bulk states.
Moreover, one can show numerically that the degree of
localization near the chain ends  increases with increasing 
chain-lead coupling.
Secondly, we find that the coupling is able to  generate edge states 
in the gap even under non-topological conditions (i.e., for $t_2/t_1<1$) if 
the parameter $\tau_c$ is sufficiently strong.

\section{Quantum transport on topological edge states and
chiral disorder effects}
From the point of view of the charge transport through the SSH chain, one 
may expect, at the first sight, that the conductance  should vanish in
the range of the topological states because of their localized character.
However, the numerical calculation based on the Landauer approach 
indicates a non-vanishing transmission coefficient at energies 
in the middle of the gap, where the topological states are 
located (see Fig.8). An exact analytical calculation is also accessible 
for the energy $E=0$, and the result Eq.(22) indicates the same, 
and also the vanishing of the conductance in the limit of long chains.

Obviously,  the coupling to  external leads should affect mostly
the conductance  of the states localized at the edges of the system, 
and this is indeed visible in Fig.8. 
One notices that for small coupling $\tau_c$ the two peaks 
in  the middle of the spectrum  are  well separated, but with increasing 
$\tau_c$ they overlap and ends up with a fast decay of the transmission.

With the effective Hamiltonian (16) at hand,  one may calculate the
Green function $G(z)=(z-H_{eff})^{-1}$ that enters the expression of the 
transmission coefficient in  the Landauer-B{\"u}ttiker formalism.
The Fermi energy in the leads being fixed at $E_F=0$,  the whole energy 
spectrum of the SSH chain can be scanned by applying a gate potential 
$V_{gate}$, which is varied continuously \cite{Note-Vg}. Assuming that 
the leads are attached to the first and last site $n=1,N$
of the SSH chain, the transmission coefficient is given in units $e^2/h$
 by \cite{Caroli}:
\begin{equation}
T(V_{gate},\tau_c)=\frac{4\tau_c^4}{t_L^2}
|G_{1N}(V_g)|^2 sin^{2}k
=\frac{4\tau_c^4}{t_L^2}|<1|(V_g-H_{eff})^{-1}|N>|^2 sin^{2}k,
\end{equation}
where $sin k$ comes from the density of states in the semi-infinite 
leads. We remind that $E_F=0$ corresponds to $sink=1$.

Usually, Eq.(18) is evaluated numerically  by scanning the whole 
spectrum  as it is shown in Fig.8 for a chain of length $N=8$. 
One  remarks  that the transmission peaks on the bulk states show a robust
unitary limit ($T=1$) with respect to lead-chain coupling 
strength, while the transmission on the  topological states
is very sensitive to  this coupling.

It is notable that $T(V_{gate}=0)$, which corresponds to the electron 
transmission at the middle of the gap, can be calculated analytically exact.
\begin{figure}[htbp]
\centering
\includegraphics[scale=1.5]{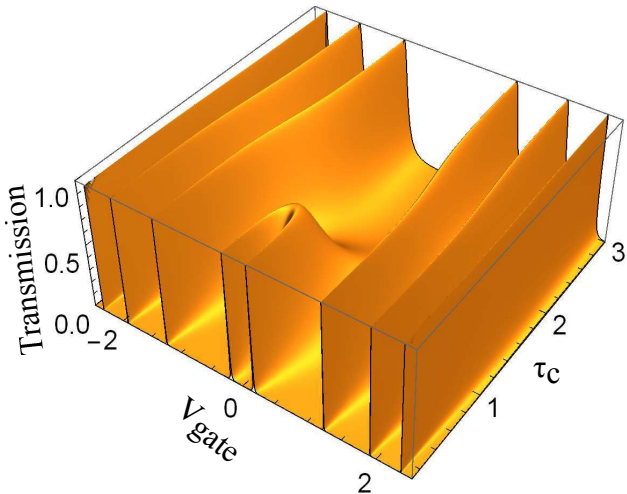}
\caption{Transmission coefficient as function of the chain-lead 
coupling $\tau_c$ for  the  SSH chain of length $N=8$. 
The lateral peaks correspond to the six bulk states, the central peaks 
are due to the topological edge states ($t_1=1, t_2=1.5$).}
\end{figure}
To this aim, besides the function $G$, one has to consider
$G^0(z)=(z-H)^{-1}$, 
of the finite (isolated) SSH chain. 
Then, the matrix element $G_{1N}$, needed in Eq.(18), is  given by the 
Dyson equations:
\begin{eqnarray}
\hskip2cm
G_{1N}=G^0_{1N}+G^0_{11}\gamma G_{1N}+G^0_{1N}\gamma G_{NN},~~~~\nonumber\\
\hskip2cm
G_{NN}=G^0_{NN}+G^0_{N1}\gamma G_{1N}+G^0_{NN}\gamma G_{NN},
\end{eqnarray}
with $\gamma=i(\tau_c^2/t_L) $ (see Eq.12).
It results immediately:
\begin{equation}
G_{1N}=G^0_{1N}/[(1-\gamma G^0_{11})(1-\gamma G^0_{NN})- 
\gamma^2 G^0_{1N} G^0_{N1}].
\end{equation}
The matrix elements of  $G^0$ can easily  be calculated analytically 
using the recipe \cite{Fonseca} for the inversion of tridiagonal matrices:
\begin{equation}
G^0_{1N}=\big(H^{-1}\big)_{1N} = 
{\frac{1}{t_2}}\big(\frac{t_2}{t_1}\big)^{N/2},
\end{equation}
and also $G^0_{11}= G^0_{NN}=0$. By using these expressions in (20), 
and next in (18), we obtain the following analytical formula for 
the transmission coefficient of the SSH finite ordered chain 
with an {\it even} 
number of sites:
\begin{equation}
T(V_{gate}=0,\tau_c)=\frac{\big(4\tau_c^{4}/{t_L}^2 t_2^2\big)(t_2/t_1)^N}{[1+
\big(\tau_c^{4}/{t_L}^2 t_2^2\big)(t_2/t_1)^N]^2}.
\end{equation}
The  transmission coefficient (22) shows an intriguing {\it non-monotonous} 
dependence on the chain-lead  coupling parameter $\tau_c$, 
which is depicted in Fig.9a (red curve). The maximum corresponds 
to a unitary transmission, the position of which can be deduced immediately as:
\begin{equation}
\tau_c^{max}=\sqrt{t_L t_2} \big(t_1/t_2\big)^{N/4}.
\end{equation}

One observes in Fig.9a that the disorder gives rise to a shift to 
the right of this maximum. This aspect
cannot be overlooked as it  proves a disorder induced increase of 
the transmission coefficient, whenever the chain-lead coupling 
is strong enough. 
\begin{figure}[htbp]
\centering
\includegraphics[angle=-90, scale=0.27]{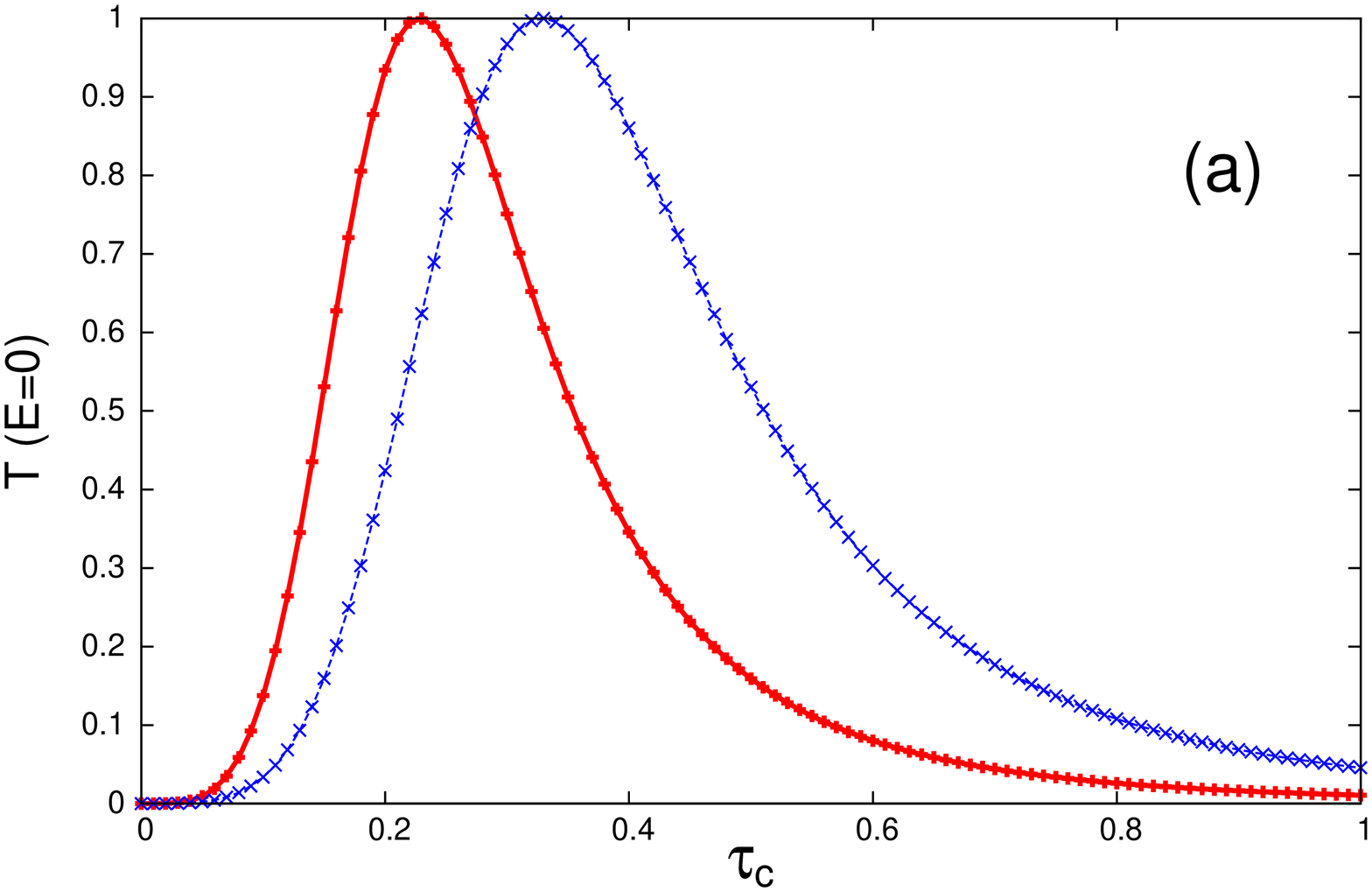}
\includegraphics[angle=-90, scale=0.27]{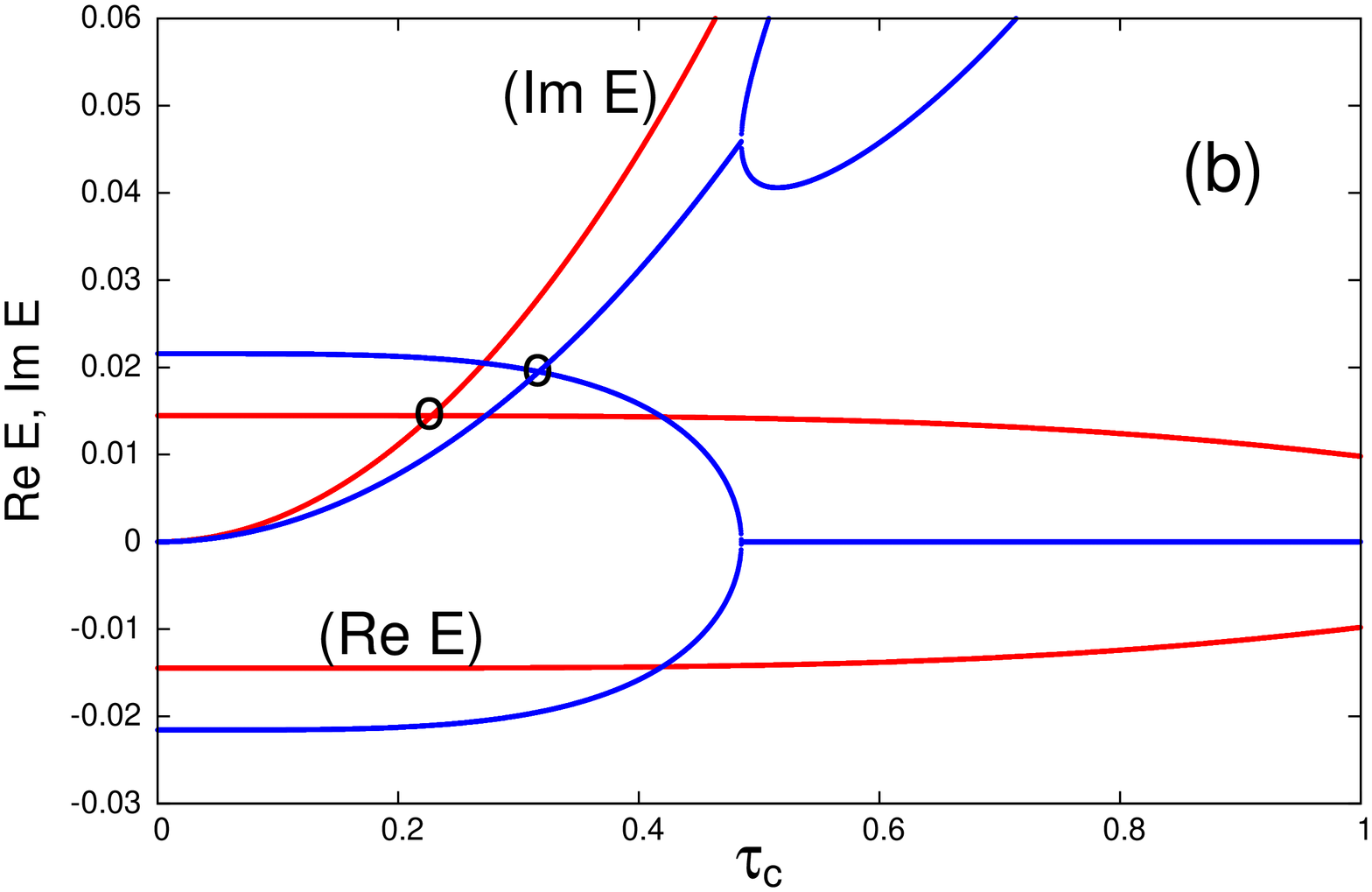}
\caption{Transmission coefficient at $E=0$ (a) and complex 
topological eigenvalues (b)
as function of the chain-lead coupling $\tau_c$ for  ordered (red) 
and  disordered (blue) SSH chain. 
In both cases, the unitary peak of $T(E)$ corresponds to the 
intersection $Re E=Im E$ in the spectrum ($N=20, t_2/t_1=1.5, W=2$).}
\end{figure}

Aiming to get   insight into the  physics behind the  non-monotonous 
behavior of the transmission coefficient, we approach  below
the question of the density of states DoS in the energy range of 
the topological states. Obviously, the  coupling to the leads gives 
rise to the broadening of the energy levels. The resulting  overlap 
of the energy tails creates a non-vanishing (and increasing with $\tau_c$)  
density of states at $E=0$, which can support the increase of the 
transmission coefficient on the left-hand side of the red curve in Fig.9a.
On the other hand, one may also speculate that  the coalescence process 
shown in Fig.6  affects the  DoS at high values of the coupling parameter.

Formally, the density of states reads:
\begin{equation}
\begin{split}
   g(E,\tau_c) &= -\frac{1}{\pi} Im \sum_i \langle\Psi_i| (E-H_{eff})^{-1}|\Psi_i\rangle \\
               &= -\frac{1}{\pi} Im \sum_i(E-Re E_i +i Im E_i)^{-1},
\end{split}
\end{equation}
however at $E=0$, only the two poles coming from the topological
levels $E_{-}$ and $E_{+}$  contribute to the density of states. 
Since the real parts  of two levels are symmetric about $Re E=0$ 
($ReE_{+}=-ReE_{-}=\Delta/2$), and the imaginary parts are 
the same ($ImE_{-}=ImE_{+}=\Gamma$), one obtains:
\begin{equation}
	g(E=0, \tau_c)= \frac{2}{\pi}\frac{\Gamma(\tau_c)}
{\Delta^2(\tau_c)/4+\Gamma^2(\tau_c)}. 
\end{equation}
Let us  calculate the real and imaginary part of $E_{+}$ and $E_{-}$ 
and represent them  simultaneously in Fig.9b. 
We note that $|Re E_{\pm}|$ decreases with $\tau_c$, while the 
imaginary part behaves oppositely. As a result, as long as $\Delta/2>\Gamma$, 
the density of states (25)   increases with $\tau_c$, but decreases 
in the opposite case $\Delta/2<\Gamma$. So, the maximum of the 
density of states is reached for that value of $\tau_c$  where 
the real and imaginary part of the topological eigenvalues 
intersect:
\begin{equation}
\Delta(\tau_c)/2=\Gamma(\tau_c) .
\end{equation}
The first remark is that the value of $\tau_c$
corresponding to the intersection coincides with the value 
of $\tau_c$ where the transmission coefficient reaches
its maximum in Fig.9a. This occurs both for the ordered and disordered chain 
(red and blue curves, respectively).
The second remark is that the curves
corresponding to the   disordered chain, intersect at a higher value 
of the coupling than  for the ordered chain. 
The implications of this last observation will be discussed later
in the context of disorder effects.

According to  Fig.6, for larger couplings, close to  the coalescence, 
one has $\Delta(\tau_c) \approx 0$, while $\Gamma(\tau_c)$ keeps 
increasing with $\tau_c$, fact that implies a small lifetime 
$\tau=\hbar/\Gamma$ of the topological states. In this range, 
the density of states (25)  becomes:
\begin{equation}
	g(E=0, \tau_c)= 2/\pi\Gamma=2 \tau/\pi \hbar.
\end{equation}
It turns out that the continuous decrease of the lifetime gives rise to
the continuous depletion of the density of states, which in turn
yields the decrease of the transmission coefficient also beyond the 
coalescence point.

The  conclusion of the above discussion is that the non-monotonous 
behavior of $T(V_{gate}=0,\tau_c)$ follows  the behavior of the  
density of states in the middle of the gap. 

As we have already discussed, the non-Hermitian 
Hamiltonian (12), which describes the  transport  problem,
does not exhibit  exceptional points in the case of ordered even chains.
However, let us consider also  an associated Hamiltonian:
\begin{equation}
H_{PT}=H+\frac{i\tau_c^2}{t_L}(c_1^{\dag}c_1-c_N^{\dag}c_N),
\end{equation}
which shows $\mathcal{PT}$-symmetry and an exceptional point, 
which depends on $\tau_c$. It is  remarkably interesting that the 
position of the EP shown by the spectrum  of (28) coincides
with the transmission peak position at $\tau_c=\tau_c^{max}$.
This can easily be proved analytically (by looking for the poles of
the Green function $G(z)=(z-H_{PT})^{-1}$) and checked numerically 
by calculating the energy spectrum. Such spectral similarities 
between the two Hamiltonians of the ordered chains 
are studied also in \cite{Gorbatsevich}. 

More insight on the manifestation of the topological states 
in open systems can be obtained  by examining other relevant quantities,
like  the {\it dwell time}  and  {\it reflected flux delay}. 

The dwell time $\tau_D$ measures the time spent in the system 
by the electron injected from the lead. According to the definition 
\cite{Buettiker}, the calculation of the dwell time pretends the 
knowledge of the probability to find the incident particle inside 
the barrier, which next has to  be divided by the incident current. 
For  our specific problem, the barrier consists of the SSH chain, 
and the quantities of interest are to be calculated in the energy 
range of the topological states.

The Lipmann-Schwinger formalism provides the scattering wave 
function $|\Psi^{scatt}\rangle$, which we project on all sites $n$
of the chain. Then, the probability to find the particle inside 
the chain is  $\sum_n |\langle n|\Psi^{scatt}\rangle|^2$, and is given by 
the following expression, which we use to calculate  $\tau_D$ numerically:
\begin{equation}
\sum_n |\langle n|\Psi^{scatt} \rangle|^2=
\tau_c^2\sum_n |\langle n| H_{eff}^{-1} |1 \rangle|^2.
\end{equation}
(Notice that the above formula assumes the electron to be injected 
at the site $n=1$. The technicalities related to the deduction of 
(29) can be found in \cite{ONA}, and are no more repeated here.)
\begin{figure}[htbp]
\centering
\includegraphics[angle=-90, scale=0.27]{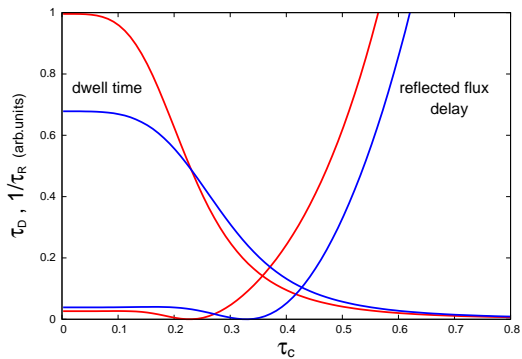}
\caption{Dwell time  and  inverse of the reflected flux delay  
as function of the coupling parameter $\tau_c$  for ordered 
(red curves) and disordered (blue curves) SSH chain in the 
topological phase ($t_2/t_1=1.5, N=20, W=2$).
The dwell time is scaled to have the maximum at $\tau_D=1$.}
\end{figure}

In contrast to the transmission coefficient, the dwell time  plotted 
in Fig.10 exhibits a monotonous decrease with the coupling parameter, 
with a significant drop about $\tau_c=\tau_c^{max}$.
One knows \cite{Winful} that $\tau_D$ cannot distinguish between the
transmitted and reflected flux, such that one cannot say  a priori 
whether  the decay comes from an increasing transmission or increasing 
reflection.

In the range of  low coupling, the answer to the above question is 
simple, as the increasing  transmission (observable in Fig.9a 
for $\tau_c<\tau_c^{max}$) obviously contributes to reducing the 
time $\tau_D$ spent by the charge inside the system. On the other hand, 
for strong coupling, we know (again from Fig.9a)  that the transmission 
falls down drastically, such that the  prevailing process  in this 
range should be the reflection.

The above  conclusion is reinforced by the behavior of the 
reflected flux delay defined as $\tau_R=\tau_D/(1-T^2)$.
It is to remind that a small $\tau_R$ evidentiates a strong
reflection process.
Then, by observing the behavior of $1/\tau_R$ in Fig.10, one notices 
indeed a strong reflection in the range of large coupling $\tau_c$.
One may also observe that the zero of the reflected flux delay 
coincides with the unitary maximum of the transmission coefficient 
in Fig.9a. 

In what follows, we examine the combined effect of the chiral disorder 
and  chain-lead coupling on the transport properties of the  SSH chain.
Using (18), the transmission coefficient can be calculated  for
any given disorder configuration. 
Fig.9a exhibits the transmission coefficient as function of the coupling
for both the  ordered chain (red curve)  and  a disordered 
chain with $W=2$ (blue curve).

The comparison  of the two curves highlights a surprising behavior, 
namely:   for weak coupling (small  $\tau_c$), the disorder effect
consists in reducing the transmission, however, in the range of  
strong coupling, the effect is opposite. One notices that the 
significantly different  behaviors are associated to the shift 
with  disorder of the transmission peak. The shift  infers  that 
the condition (26), telling where the maximum occurs,
is satisfied at higher values of $\tau_c$ in the disordered case. 
This is indeed numerically  confirmed in Fig.9b by the intersection 
of the blue curves, which describe the real and imaginary part 
of the topological eigenvalues in the disordered case.  
The shift of the transmission peak can be  understood by  replacing 
the hopping  parameter $t_2$ in the expression (23) for $\tau_c^{max}$
with a smaller effective  $t_2^{eff}$ acting in the disordered case. 
We  estimate below this effective parameter in the case of  weak disorder.

Let us remind that the chiral disorder is introduced   by turning 
$t_2$ in each cell  of the ordered chain (indexed by $i=1,...,N_c$)
into a random variable $\{t_2^{(i)}\}$, which is equally distributed 
in the range $[-\frac{W}{2},\frac{W}{2}]$. A constraint on the mean value 
is imposed for each disorder configuration, namely:
$\frac{1}{N_c}\sum_{i} t_2^{(i)}= t_2$. With  $t_2^{(i)}=t_2+\delta_i$,
the constraint becomes $\sum_i \delta_i=0$.
Then, any quantity of the type $t_2^{N_c}$, which appears in formulas
describing the ordered chain (as, for instance, in (21)), should be 
replaced by  the configurational average $<\Pi_{i=1}^{N_c} t_2^{(i)}>$.
For weak disorder, keeping only the terms up to the second power in $\delta$, 
one obtains:
\begin{equation}
<\Pi_{i=1}^{N_c} t_2^{(i)}>\approx t_2^{N_c}+t_2^{N_c-2}\sum_i\sum_{j<i}
<\delta_i \delta_j> .
\end{equation}
The fluctuations $<\delta_i \delta_j>$ are easily estimated from:
\begin{equation}
<\big(\sum_i\delta_i\big)^2>=\sum_i<\delta_i^2>+
2\sum_i\sum_{j<i}<\delta_i \delta_j>=0,
\end{equation}
where, for the specific  disorder distribution, the variance is 
known to be  $<\delta_i^2>=\frac{W^2}{12}$. It results that, for 
weak disorder,
\begin{equation}
<\Pi_{i=1}^{N_c} t_2^{(i)}>= t_2^{N_c}\big(1-\frac{N_c}{24}\frac{W^2}{t_2^2}\big),
\end{equation}
proving that the disorder induced effective hopping parameter
$t_2^{eff}$ is  smaller than $t_2$:
\begin{equation}
t_2^{eff}=t_2\sqrt[N_c] {1-\frac{N_c}{24}\frac{W^2}{t_2^2}}. 
\end{equation}
This result is responsible for the shift  of the  transmittance 
peak in Fig.9a, but also for the disordered increased splitting 
$E_+ -E_-$, which can be noticed  in Fig.9b. Obviously, the longer 
the chain, the smaller the disorder for which (33) is valid.

A full description, for any disorder, of the transmission 
on the topological states in the parameter  space $\{W,\tau_c\}$ 
is given in Fig.11. The two cases shown in Fig.9a for $W=0,2$ 
can be identified immediately. The figure brings into attention 
the position of the transmission peak (red region), which  evolves 
versus higher values of the coupling as the disorder is increased. 
This occurs, however, only up to $W\approx3$. Above this value, 
the topological states are mixed up with  disordered states coming 
from the bands, so that the topological phase is destroyed.
\begin{figure}[htbp]
\centering
\includegraphics[scale=0.8]{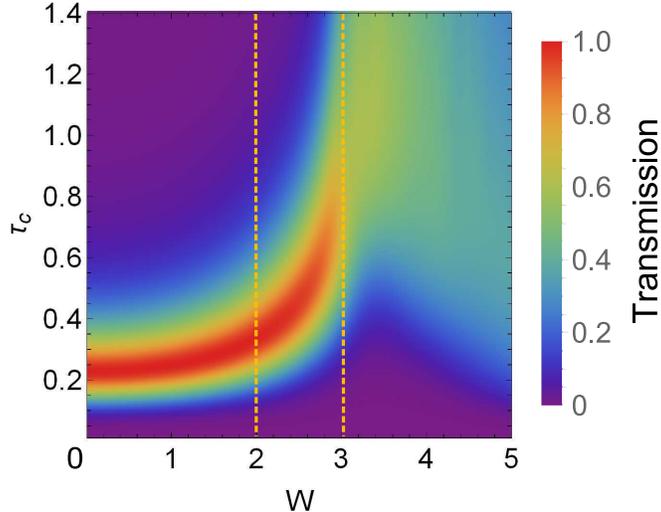}
\caption{
Transmission coefficient at $V_{gate}=0$ in the topological phase 
$t_2/t_1=1.5$  as a function of the
disorder strength $W$ and coupling parameter $\tau_c$, 
calculated by averaging over 1000 
disorder configurations. The dashed line at $W=2$ corresponds to the 
disordered case in Fig.9a. The  dashed line at $W=3$ represents the border
above which the bulk states coming from the two bands mix up
with the topological states.
The number of sites is $N=20$.}
\end{figure}

For fixed values of the lead-chain coupling,  different regimes can be 
observed in Fig.11. At very small and large values of $\tau_c$ 
the transmission  coefficient vanishes. However, two types of 
disorder dependence can be detected in the range of interest about 
the transmission peak. As examples, we observe that at   $\tau_c=0.2$, 
the transmission decreases with the disorder strength $W$, however,
at $\tau_c=0.4$ the transmission is {\it enhanced} in the presence of 
disorder. The enhancement occurs only for sufficiently large 
$\tau_c$, the effect being apparent in Fig.11 for $\tau_c \gtrsim 0.3$.

To understand the above effect, one has to keep in mind that, being 
localized near the ends of the chain, the topological states conduct 
only inasmuch as they penetrate the middle of the chain. Hence, 
the  disorder-driven increase of the transmission should be accompanied 
by a corresponding change in the local density of states along the chain.
The expectation is confirmed by the numerical calculation of the LDoS:
\begin{equation}
g_n(E=0)=-(1/\pi) Im\langle n|H_{eff}^{-1}n\rangle.
\end{equation}
Indeed, in Fig.12 we observe that the local density of states 
in the middle of the chain is sensible higher in the disordered 
case (green and blue curves) than in the ordered one (red curve).
\begin{figure}[htbp]
\centering
\includegraphics[angle=-90, scale=0.3]{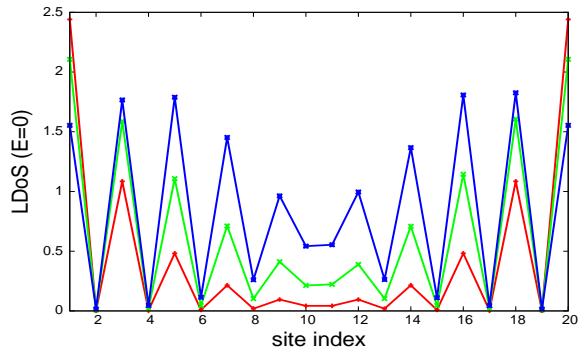}
\caption{Local density of states at $E=0$ in the ordered and
disordered  case, corresponding to $W=0$ (red), $W=2$ (green) 
and $W=2.5$ (blue). Note that the disorder increases the  density
of states in the middle of the chain. In the disordered case, 
one averages over 5000 configurations ($t_2/t_1=1.5,\tau_c=0.5$).}
\end{figure}

The response to the chiral disorder shown by the open SSH chain 
in the case of strong coupling represents a signature   of 
the {\it disorder-driven} electron conductance in 1D systems.
This effect was already met in two-dimensional systems (2D) like 
topological Anderson insulators (TAI) \cite{Jain}, or in the situation 
when the edge states coexist with bulk states and become functionalized 
by disorder \cite{Baum, Bogdan}. The possibility of finding the  
TAI phase in disordered wires, and also in disordered gain-loss 1D systems 
is discussed in \cite{Meier,Luo}.

We conclude the chapter by investigating the disorder effect on the 
dwell time and  reflected flux delay, the result being shown in Fig.10 
(blue curves).  The first notable effect is the reduced value  of 
$\tau_D$ in comparison with the ordered case, visible in the range 
of low coupling. As the transmission is also reduced in this range, 
the small dwell time comes from the disorder increased reflectivity.

In what concerns the reflected flux delay, the range of interest is 
$\tau_c>\tau_c^{max}$, where the smaller values of $1/\tau_R$ 
(compared to the ordered case) mean a lower reflectivity and, 
automatically, a higher transmission caused by disorder. 

It turns out  that the response to  disorder of the local density of 
states, dwell time and   reflected flux delay endorse the behavior
of the transmission coefficient in the presence of the chiral disorder
for the entire range of the lead-chain coupling.

\section{Summary and conclusions}
The paper has two main objectives: 
to  observe the fate of the topological edge states, 
pre-existing in the finite SSH chain, when the system is opened  by attaching
semi-infinite leads, and  to find the effects of the non-diagonal
(chiral-type) disorder and of the coupling strength on the spectral 
and transport properties of the SSH chain in the topological phase.

The physics  of the SSH chain in contact with the leads is controlled by 
three  parameters: the ratio $t_2/t_1$ (which dictates the 
topological/non-topological regime in   the finite SSH chain), the 
chain-lead coupling $\tau_c$ (which plays the role of the non-Hermitian 
parameter),  and the disorder strength $W$.

Two topological edge states, energetically located in the middle 
of the gap, arise only in the  finite SSH chains with  an even number 
of sites, which show inversion symmetry. They appear under the  condition 
of a sufficiently large ratio $t_2/t_1$, and have a finite lifetime 
coming from the non-Hermitian term in the Hamiltonian. 

The flow of the topological eigenvalues in the complex plane is 
controlled by the coupling $\tau_c$, and it is qualitatively influenced 
by the chiral disorder (see Fig.6). For the ordered chain, with 
increasing coupling, the real parts merge continuously versus $Re E=0$. 
On the contrary, the coalescence of the real parts, indicating an 
exceptional point, occurs in the disordered case, accompanied by 
the vanishing of the phase rigidity (Fig.5). In the finite SSH chain, 
the chiral disorder delocalizes the topological edge states. 
This is proved analytically by calculating the penetration length 
$\lambda$ given by Eq.(11).

Zero modes of edge-type may  appear in the spectrum of  the open 
chain even under the non-topological condition $t_2/t_1<1$.
The occurrence of these modes, evident in Fig.7e, is conditioned by a strong 
coupling to the leads, which affects  significantly the ends of the chain.

The electron transmission $T(V_{gate},\tau_c)$ of the SSH open system 
is calculated numerically using the Landauer-Buettiker formalism, 
and it makes evident in Fig.8 the transport on the topological and bulk states.

Remarkably,  an exact analytical expression Eq.(22) of the transmission 
coefficient  can be obtained  for $V_{gate}=0$, i.e., in the range of 
interest for the  topological states.  
The transmission  shows a non-monotonous dependence on the chain-lead 
coupling, and exhibits an unitary maximum, whose position depends on
$t_2/t_1$ and the chain length $N$,  and  is shifted by disorder (Fig.9).
This behavior is explained in terms of the density of states, and it is 
supported by the calculation of the dwell time  and reflected flux delay.

The  increasing strength of the chiral disorder gives rise to an
enhancement of the transmission coefficient, if the chain-lead coupling
is sufficiently strong. This is a new manifestation, this time in a 1D
system, of the disorder induced conductance in topological insulators.  

In closing, the paper  discusses the interplay between topology,
chiral disorder and non-Hermiticity, and points out a number of interesting 
spectral and transport properties, which are the result of the 
combined effect of the disorder strength and the intensity of the 
chain-lead coupling.

In view of an eventual experimental testing of the findings,
one has to mention that disordered chiral symmetric wires
were already synthesized \cite{Meier}. The remaining question is the
engineering of the broken PT-symmetry corresponding to the
non-Hermitian  Hamiltonian (13) with imaginary potential at the 
ends of the chain. It is encouraging that Song et al \cite{Song} proved 
very recently  that  non-Hermitian Hamiltonians can be implemented 
by using  wave guides techniques.

\section{Acknowledgments}
We acknowledge the financial support from Romanian Core Research 
Programme PN-21N.  The authors  thank Marian Nita and Andrei Manolescu
for very useful discussions.


\begin{thebibliography}{40}
\bibitem{Note-Vg} The gate potential applied on the SSH chain 
is simulated by introducing a diagonal term 
$V_{gate}\sum_i c^{\dag} c_i$ in the Hamiltonian $H$.
\end{thebibliography}
\end{document}